\documentclass{elsart}
\usepackage{graphicx}
\usepackage{amsmath}
\usepackage{natbib}
\begin{document}
  \begin{frontmatter}
    \title{The axial ratio of hcp iron at the conditions of the Earth's inner
      core}
    \author[ucl_pa]{C. M. S. Gannarelli}
    \author[ucl_pa,ucl_es]{D. Alf\`{e}}
    \author[ucl_pa]{M. J. Gillan\corauthref{mjg}}
    \corauth[mjg]{Corresponding Author.
      Tel: +44 (0)20 7679 7049;
      Fax: +44 (0)20 7679 1360}
    \ead{m.gillan@ucl.ac.uk}
    \ead[url]{http://www.cmmp.ucl.ac.uk/\~{}mjg}
    \address[ucl_pa]{%
      Physics and Astronomy Department,
      University College London,\\
      Gower Street,
      London WC1E 6BT,
      United Kingdom}
    \address[ucl_es]{%
      Department of Earth Sciences,
      University College London,\\
      Gower Street,
      London WC1E 6BT,
      United Kingdom}
    \begin{abstract}
      We present \textit{ab initio} calculations of the high-temperature
      axial \(c/a\) ratio of hexagonal-close-packed (hcp) iron at Earth's core
      pressures, in order to help interpret the observed seismic anisotropy of
      the inner core. The calculations are based on density functional theory,
      which is known to predict the properties of high-pressure iron
      with good accuracy. The temperature dependence of \(c/a\) is determined
      by minimising the Helmholtz free energy at fixed volume and temperature,
      with thermal contributions due to lattice vibrations calculated using
      harmonic theory. Anharmonic corrections to the harmonic predictions are
      estimated from calculations of the thermal average stress obtained from
      \textit{ab initio} molecular dynamics simulations of hcp iron at the
      conditions of the inner core. We find a very gradual increase of axial
      ratio with temperature. This increase is much smaller than found in
      earlier calculations, but is in reasonable agreement with recent
      high-pressure, high-temperature diffraction measurements. This result
      casts doubt on an earlier interpretation of the seismic anisotropy of the
      inner core.
    \end{abstract}
  \end{frontmatter}

  \section{Introduction}

  In the last few years, there have been a number of \textit{ab initio} studies
  of iron at high temperatures and pressures, in an effort to understand the
  properties of the Earth's solid inner core
  (\cite{scs:1994,sc:1995,smw:1996,wsc:1996,swc:1997,vwkgp:1997,ssc:1999,
    vbagp:2000,baj:2003,lbcst:2000,apg:2001,sscg:2001,ssc:2002,vagwbp:2003}),
  including its elastic properties. A controversial issue, addressed by
  some of this work, concerns the elastic anisotropy of the inner core
  (\cite{creager:1992,tromp:1993}), i.e. the fact that compressional waves are
  observed to traverse the core region some 3--4\% faster along the Earth's
  rotational axis than in the equatorial plane.  It is widely assumed that
  the phase of Fe in the inner core is hexagonal close packed (hcp), and it
  has become clear that to understand the elastic properties it is necessary to
  know how the axial ratio \(c/a\) depends on temperature and pressure. In
  order to provide information about this, we present here \textit{ab initio}
  calculations of the axial ratio of hcp iron over a range of pressures and
  temperatures relevant to the inner core.

  \textit{Ab initio} calculations on pure crystalline iron are very relevant
  for understanding the inner core, even though the core is known to contain
  light impurities (the leading candidates are O, Si and S), and is believed
  to contain Ni as well as Fe (\cite{poirier:1994}). Recent estimates
  (\cite{agp:2002}) suggest that the fraction of
  light impurities in the inner core is approximately 8.5~mol\%, and this low
  fraction will probably not change its properties greatly. The Ni content will
  also probably make only a small difference, since the electronic structures
  of Ni and Fe are so similar. The determination of the properties of pure Fe
  is thus essential as a starting point for further refinements. The common
  assumption that Fe in the inner core is in the hcp structure has recently
  been challenged (\cite{vagwbp:2003,baj:2003}), and it is
  possible that the crystal structure is different in different parts of the
  core (\cite{sh:1998,id:2002,bt:2003}). However, in order to make
  progress, it is clearly essential to understand the properties of the leading
  candidates, of which hcp is certainly one.  It is well established that
  density functional theory (DFT) (\cite{hk:1964,ks:1965,jg:1989}) gives a
  rather accurate description of Fe and other transition metals, both under
  ambient conditions, and under high pressures and temperatures.  This is known
  from comparisons with experiment for a wide range of properties, including
  the equilibrium lattice parameter under ambient conditions, elastic
  constants, magnetic properties, phonon frequencies
  (\cite{scs:1994,smw:1996,vwkgp:1997,vbagp:2000}); the pressure-volume
  relationship at high pressures, the phonon density of states and the Hugoniot
  (\cite{wsc:1996,mao_ea:2001,apg:2001}).

  A number of groups have used DFT calculations to investigate the elastic
  constants of hcp Fe under inner-core conditions.  The early calculations were
  performed at zero temperature (\cite{sc:1995,ssc:1999}), and led to the
  suggestion that elastic
  anisotropy could be explained by a preferential alignment of crystalline
  \(c\)-axes with the Earth's rotational axis.  However, more recent work has
  indicated a strong temperature dependence of the elastic constants
  (\cite{sscg:2001,ssc:2002}), including
  a crossing of the \(c_{11}\) and \(c_{33}\) constants at approximately
  1500~K, which would result in a reversal of the necessary crystalline
  alignment.  It appears that the strong temperature dependence of the elastic
  constants is due in part to a rapid increase of the axial ratio with
  increasing temperature. However, this work can be questioned, because it is
  based upon a statistical mechanical approximation known as the
  particle-in-cell (PIC) model
  (\cite{hcb:1954,hr:1970,h_ea:1970,rh:1973,wc:1975,c_ea:1990}), which may not
  be reliable. In addition,
  independent PIC calculations have failed to reproduce their results
  (\cite{gag:2003}). Very recently, experimental work (\cite{mssmsh:2004}) has
  shown that this
  strong \(c/a\) variation with temperature is not observed at pressures up to
  160~GPa. Our aim
  in this paper is to perform calculations of the high temperature axial ratio,
  which are, as far as possible, free of statistical mechanical approximations.
  
  The calculation of phonon dispersion relations using DFT is nowadays
  completely routine, and this makes it possible to obtain the Helmholtz free
  energy \(F\) of a high-temperature crystal without any statistical mechanical
  approximations, provided that anharmonicity can be ignored. Phonon dispersion
  relations for Fe in both the magnetic body-centred cubic and the hcp
  structures have been reported by several groups (\cite{vbagp:2000}). The
  equilibrium value of \(c/a\) can then be obtained for given \(V\) and \(T\)
  by minimising \(F\) with respect to \(c/a\). Of course, at temperatures near
  the melting point, it is not obvious that anharmonic corrections to \(F\) can
  be neglected. However, direct DFT molecular dynamics simulations allow the
  calculation of the thermal average stress tensor (\cite{obp:2001}), so that
  the equilibrium \(c/a\) can be
  found by requiring that the stress tensor be isotropic; such direct
  simulations completely include anharmonicity. The DFT calculation of phonon
  frequencies is straightforward and relatively inexpensive, but
  \textit{ab initio} molecular dynamics demands much larger computational
  resources. Our strategy in this work is therefore to base most of the
  calculations on the harmonic approximation, but to use a small number of
  molecular dynamics simulations to estimate anharmonic corrections to the
  equilibrium \(c/a\).
  
  The rest of this paper is organised as follows: In Sec.~\ref{sec:theory} we
  present the technical background to our electronic structure calculations,
  and describe the statistical mechanical principles of
  our harmonic and molecular-dynamics calculations. In Sec.~\ref{sec:harm} we
  describe our harmonic calculations, including the technical tests we
  performed to ensure their accuracy. In Sec.~\ref{sec:md} we  present our
  \textit{ab initio} molecular-dynamics results. Discussion of our results for
  the temperature and pressure dependence of the axial ratio follows in
  Sec.~\ref{sec:conc}.

  \section{Techniques}
  \label{sec:theory}
  The principal quantity in this work is the \textit{ab initio} Helmholtz
  free energy \(F(V,q,T)\) as a function of the volume \(V\), axial ratio
  \(q=c/a\) and temperature \(T\).  Since the temperatures of interest here are
  far above the Debye temperature, \(F\) is given by the standard
  expression of classical statistical mechanics
  \begin{equation}
    F(V,q,T)=-k_{\rm B}T \ln \left\{\frac{1}{\Lambda^{3N}}\int d\mathbf{r}_1
    \ldots d\mathbf{r}_N
    \exp[-U_{\rm AI}(\mathbf{r}_1 \ldots \mathbf{r}_N;T_{\rm el})
    /k_{\rm B}T]\right\}
  \end{equation}
  where \(T\) is the temperature and \(\Lambda\) is the thermal wavelength. The
  quantity \(U_{\rm AI}(\mathbf{r}_1\ldots\mathbf{r}_N;T_{\rm el})\) is the
  \textit {ab initio} Helmholtz free energy
  of the system when the atoms are fixed at the positions \(\{\mathbf{r}_i\}\);
  \(U_{\rm AI}\) is a free energy because it includes thermal excitation of the
  electrons, at electronic temperature \(T_{\rm el}\), as described by
  finite temperature density functional theory
  (\cite{mermin:1965}). Since we are interested in the case of full thermal
  equilibrium, we have \(T_{\rm el}=T\). In the present work, the DFT
  calculation of
  \(U_{\rm AI}\) is made using the exchange-correlation energy given by the
  PW91 generalised gradient approximation (\cite{wp:1991}). The implementation
  of DFT is the projector augmented wave (PAW) scheme
  (\cite{blochl:1994,kj:1999}) with core radii, augmentation charge radii, etc.
  as reported in (\cite{akg:2000}). As in our previous work, atomic states up
  to and including \(3p\) are treated as core states, but the high-pressure
  response of \(3s\) and \(3p\) states is included via an empirical pair
  potential. This approximation is tested using spot-checks with \(3p\)
  electrons included in the valance set. All calculations were performed using
  the VASP code (\cite{kf_01:1996,kf_02:1996}).

  In practice, the \textit{ab initio} Helmholtz free energy is separated into
  perfect lattice, harmonic and anharmonic contributions:
  \begin{equation}
    F=F_{\rm perf}+F_{\rm harm}+F_{\rm anharm}\;.
  \end{equation}

  Here \(F_{\rm perf}=U_{\rm AI}(\mathbf{R}_1\ldots\mathbf{R}_N;T_{\rm el})\)
  is the
  \textit{ab initio} free energy with all atoms fixed on their perfect-lattice
  positions \(\{\mathbf{R}_i\}\). The harmonic contribution \(F_{\rm harm}\)
  is given by (\cite{apg:2001})
  \begin{equation}
    F_{\rm harm}=3k_{\rm B}T \ln \frac{\hbar\bar{\omega}}{k_{\rm B}T}\;,
  \end{equation}
  with \(\bar{\omega}\) the geometric mean phonon frequency:
  \begin{equation}
    \ln\bar{\omega}=\frac{1}{N_{\mathbf{k}s}}\sum_{\mathbf{k}s}
      \ln\omega_{\mathbf{k}s}\;,
  \end{equation}
  where \(N_{\mathbf{k}s}\) is the total number of \(k\)-points and
  phonon-branches, the
  sum being taken over the first Brillouin zone. \(\bar{\omega}\) is a function
  of \(V\), \(q\) and \(T_{\rm el}\). As usual, the phonon frequencies
  \(\omega_{\mathbf{k}s}\) at each wavevector \(\mathbf{k}\) are obtained by
  diagonalising the dynamical matrix, whose elements are defined in terms of
  the force-constant matrix \(\Phi_{ls\alpha,l't\beta}\) as:
  \begin{equation}
    D_{s\alpha,t\beta}(\mathbf{k})=\frac{1}{M}\sum_{l'}\Phi_{ls\alpha,l't\beta}
    \exp\left[i\mathbf{k}.(\mathbf{R}^0_{l't}-\mathbf{R}^0_{ls})\right]\;,
  \end{equation}
  where \(M\) is the mass of each atom.
  Here \(\mathbf{R}^0_{ls}\) is the equilibrium position of the \(s\)th atom in
  the \(l\)th primitive cell. Calculation of the force-constant matrix \(\Phi\)
  is performed by the small-displacement method
  (\cite{kfh:1995,apg:2001,alfe_phon}), in which each atom in the primitive
  cell is given a small displacement, and DFT is used to calculate the
  resulting force on every atom in a large repeating cell. The calculation of
  the dynamical matrix and hence the mean frequency \(\bar\omega\) must be
  taken to convergence with respect to the size of the repeating cell. For the
  hcp structure, the entire \(\Phi\) matrix is calculated by performing only
  two independent displacements, as explained in \cite{apg:2001}. Within the
  harmonic approximation, the equilibrium \(c/a\) at given atomic volume and
  temperature is determined by calculating \(F_{\rm perf}(q)\) and
  \(F_{\rm harm}(q)\) at a series of \(q\) values, and by minimising the
  quantity \(F_{\rm perf}(q)+F_{\rm harm}(q)\) numerically.

  Our estimates of anharmonic corrections to the equilibrium \(c/a\) are
  based on \textit{ab initio} molecular dynamics simulations. These are
  performed in the canonical ensemble, using an Andersen thermostat
  (\cite{andersen:1980}). The
  determination of the equilibrium \(c/a\) is based on a calculation of the
  time-averaged stress tensor \(\sigma_{\alpha\beta}\). The system is in 
  hydrostatic equilibrium with respect to variation of \(q\) when
  \(\sigma_{33}-\sigma_{11}=0\). In comparing
  with harmonic predictions, it is useful to note that the components of the
  thermal average stress tensor are related to the \textit{ab initio}
  harmonic free energy by
  \begin{equation}
    <\sigma_{\alpha\beta}>=\frac{1}{V_0}\lim_{\varepsilon_{\alpha\beta}
      \rightarrow 0}\left(\frac{\partial F}{\partial\varepsilon_{\alpha\beta}}
    \right)_T\;,
  \end{equation}
  where \(\varepsilon_{\alpha \beta}\) is the infinitesimal strain tensor and
  \(V_0\) is the volume of the system before application of the strain.
  In particular, for our constant volume calculations of \(F(V,q,T)\), we can
  write
  \begin{equation}
    \label{eq:stress}
    <\sigma_{33}-\sigma_{11}>=\frac{3q}{2V}
    \left(\frac{\partial F}{\partial q}\right)_{V,T}\;.
  \end{equation}
  This means that the stress component \(<\sigma_{33}-\sigma_{11}>\) within the
  harmonic approximation can be obtained by taking the derivative with respect
  to \(q\) of the free energy \(F_{\rm perf}(q)+F_{\rm harm}(q)\).

  \section{Harmonic calculation of the equilibrium axial ratio}
  \label{sec:harm}
  In order to determine the equilibrium axial ratio in harmonic theory, the
  two quantities that it is necessary to calculate are the perfect lattice free
  energy \(F_{\rm perf}\), and the geometric mean phonon frequency
  \(\bar{\omega}\), both as function of volume, axial ratio \(c/a\) and
  electronic temperature. The calculation of these quantities is presented in
  the following subsections, and the results for the equilibrium axial ratio
  are presented in Sec.~\ref{ssec:results}.  The precision we need to
  achieve in the calculations is set by the precision with which we wish to
  determine the equilibrium axial ratio. As a guideline, we set ourselves the
  target of obtaining \(c/a\) to a precision of \(\pm 0.005\).

  \subsection{Perfect lattice free energy}
  
  DFT results for the perfect lattice free energy for \(c/a=1.6\) were reported
  earlier by \cite{apg:2001}, and were more recently reported by
  \cite{gag:2003} for \(q\) in the range 1.48--1.72. The present calculations
  are closely related to the previous ones, but we have introduced some
  refinements. We have calculated \(F_{\rm perf}\) for electronic temperatures
  in the range 1000--7000~K. For \(T_{\rm el}\) in this range, we find that an
  \(8\times8\times5\) Monkhorst-Pack \(k\)-point sampling set provides a
  precision of better than 1.5~meV~atom\(^{-1}\). The plane-wave cutoff has
  been set to achieve complete continuity in the curve \(F_{\rm perf}(q)\), and
  to ensure that at every point on this curve, the calculated energy and stress
  components are converged to within our required targets. 

\begin{figure}
    \vspace{4mm}
    \includegraphics[height=80mm]{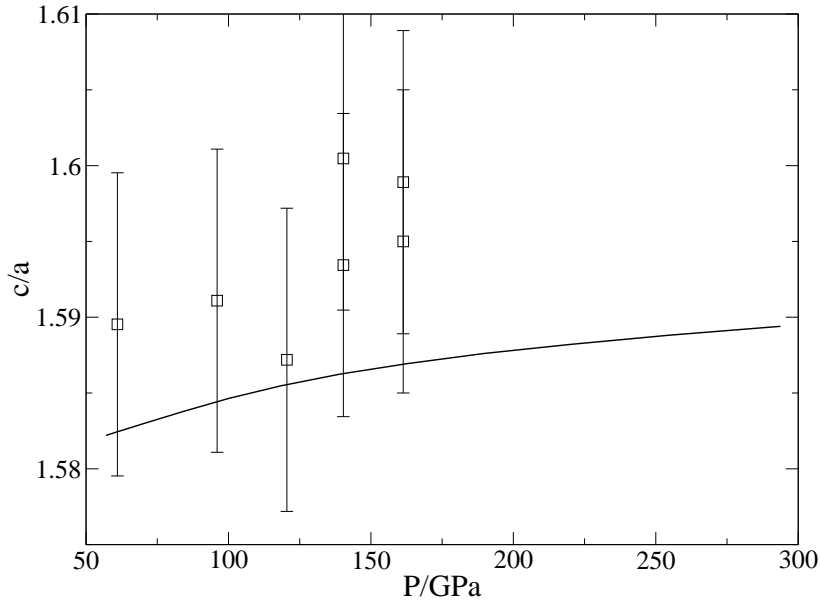}
    \caption{Calculated equilibrium axial ratio of hcp Fe at zero temperature
    as a function of pressure (solid curve), compared with diffraction results
    of \cite{mssmsh:2004} (open squares with error bars).}
    \label{ca_zero}
  \end{figure}

  For the set of \(T_{\rm el}\) mentioned above, we have calculated
  \(F_{\rm perf}\) for \(c/a\) in the range 1.54--1.70, at steps of 0.01
  and for volumes between 6.8 and 8.8~\AA\(^3\) at steps of 0.2~\AA\(^3\), as
  well as at three particular volumes of interest, 6.97, 7.50 and
  8.67~\AA\(^3\).
  A parameterisation essentially exactly fitting the data consists of a cubic
  polynomial in \(c/a\), whose coefficients are fitted to quartic polynomials
  in \(T_{\rm el}\). As a point of reference for our later discussion, we have
  used these results to calculate the equilibrium axial ratio at \(T=0\) as a
  function of volume. We have converted the results to obtain \(q\) as a
  function of pressure by using the \(P(V)\) relation from \cite{scs:1994}. In
  Fig.~\ref{ca_zero}, we compare the resulting values of \(q(P)\) with very
  recent synchrotron measurements (\cite{mssmsh:2004}) in the pressure range
  60--160~GPa. In both cases, we see a very slight increase of \(q\) with
  pressure, though the theoretical values are lower by about 0.008.

  \subsection{Harmonic calculations}

  As explained in Sec.~\ref{sec:theory}, the harmonic vibrational free
  energy is determined entirely by \(\bar{\omega}\). Values of \(\bar{\omega}\)
  must be converged with respect to \(k\)-point sampling, cell size and
  atomic displacement, and we must ensure that this convergence is achieved
  across the required range of \(c/a\). In order to achieve the required
  tolerance in the harmonic free energy, systematic errors in \(\bar{\omega}\)
  must not vary with respect to \(q\) by more than 0.1\%.

  We have designed a strategy for ensuring convergence of \(\bar{\omega}\).
  In our initial tests, we adopt a displacement of 0.01~\AA, which has
  been used in previous calculations (\cite{apg:2001}). With this displacement,
  we determine the \(k\)-point sampling needed to obtain convergence for a 
  16-atom supercell. We then study convergence with respect to cell size, the
  \(k\)-point sampling for each cell size being chosen in the light of the
  \(k\)-point sampling tests of the 16-atom system.  This set of tests tells us
  the cell size needed. To conclude the tests, we then address convergence of
  atomic displacement and \(k\)-point sampling for this cell size. All these
  tests are performed at \(T_{\rm el}=1000\)~K. This temperature is
  deliberately chosen to be much lower than the temperatures of real interest.
  We do this because the fineness of \(k\)-point sampling needed to achieve a
  given degree of convergence decreases with increasing temperature. One can
  therefore be sure that if convergence is achieved with respect to \(k\)-point
  sampling at 1000~K, it will certainly be achieved at higher temperatures.

  For the 16-atom system, \({\bar \omega}\) was calculated for Monkhorst Pack
  sets up to \(9\times9\times6\). We find that for a Monkhorst Pack
  \(5\times5\times3\) grid, \(\bar{\omega}\) is converged to within 0.01\%.
  Next, a set of supercells was chosen, such that
  the dimension along the the hexagonal axis is similar to the dimension in the
  basal plane. Sizes chosen were 16 (\(2\times2\times2\)), 36
  (\(3\times3\times2\)), 54 (\(3\times3\times3\)), 96 (\(4\times4\times3\)),
  128 (\(4\times4\times4\)) and 150 (\(5\times5\times4\)) atoms. For each cell
  size, we calculate \(\bar{\omega}\) using \(k\)-point sampling that is
  equivalently finer than the converged value of \(5\times5\times3\) in the
  16-atom system. We find that for a 54-atom system, \(\bar{\omega}\) is
  converged to within 0.5\%, however the consistency between calculations for
  different values of \(c/a\) at this cell size, is much better than our
  tolerance. Finally, \(k\)-point and displacement convergence were carried out
  on the 54-atom system. We found that for a \(3\times3\times2\) Monkhorst
  Pack grid and a displacement of 0.01~\AA, \(\bar{\omega}\) is converged to
  within approximately 1\% in absolute terms, but, again, to well within our
  tolerances for non-cancelling errors. These tests were all performed for both
  \(c/a=1.60\) and 1.70. All the harmonic results that follow were based upon
  values of \(\bar{\omega}\) calculated for these values of supercell size,
  \(k\)-point sampling and displacement.

  Calculations of \(\bar{\omega}\) were performed for atomic volumes of 6.97,
  7.50 and 8.97~\AA\(^3\), for \(c/a\) from 1.58 to 1.70 in steps of
  approximately 0.03, and for electronic temperatures of 2000, 4000 and 6000~K.
  We parameterised \(\ln \bar{\omega}(q)\) for each volume and temperature
  using a second order polynomial in \(q\), which gave an effectively exact fit
  to the results.

  \subsection{Harmonic results for equilibrium axial ratio}
  \label{ssec:results}

\begin{figure}
    \vspace{4mm}
    \includegraphics[height=80mm]{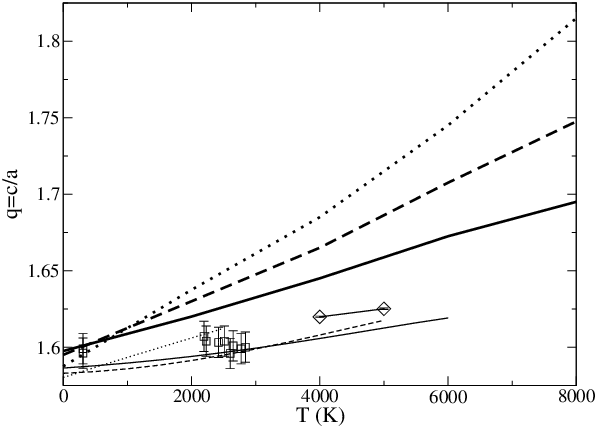}
    \caption{Calculated equilibrium axial ratio as a function of temperature
      for different volumes. For this work (light curves)
      atomic volumes are 6.97~\AA\(^3\) (solid curve), 7.50~\AA\(^3\) (dashed
      curve) and 8.67~\AA\(^3\) (dotted curve). For \cite{sscg:2001} (heavy
      curves) volumes are 6.81~\AA\(^3\) (solid curve), 7.11~\AA\(^3\) (dashed
      curve) and 7.41~\AA\(^3\) (dotted curve). Also shown are diffraction
      measurements due to \cite{mssmsh:2004} at 7.73~\AA\(^3/\mbox{atom}\)
      (open squares with error bars) and our own \textit{ab initio} MD
      calculations at 6.97~\AA\(^3\).}
    \label{ca_temp}
  \end{figure}

  Equilibrium values of \(c/a\) were obtained at 6.97, 7.50 and 8.67~\AA\(^3\)
  and temperatures of 0, 2000, 4000 and 6000~K, by analytic minimisation of the
  total harmonic free energy \(F_{\rm perf}(q)+F_{\rm harm}(q)\), using the
  polynomial parameterisations
  described above for \(\ln\bar{\omega}(V,q,T_{\rm el})\) and
  \(F_{\rm perf}(V,q,T_{\rm el})\). Results are shown in Fig.~\ref{ca_temp},
  together with the previous theoretical results of \cite{sscg:2001} and the
  very recent experimental results of \cite{mssmsh:2004}. Note that all the
  theoretical results show the variation of \(c/a\) with \(T\) at fixed volume,
  whereas the experimental results are at the fixed pressure of 161~GPa. The
  present results differ greatly from the earlier theoretical results, in that
  we find only a very moderate increase of \(c/a\) with temperature. The
  experimental increase of \(c/a\) with temperature is also far smaller than
  the predictions of \cite{sscg:2001}, and is, if anything, smaller than the
  variation that we predict. More detailed discussion will be given in
  Sec.~\ref{sec:conc}.

  \section{Molecular dynamics calculations}
  \label{sec:md}

  The determination of the equilibrium axial ratio by \textit{ab initio}
  molecular dynamics is based on the calculation of stresses; specifically on
  the stress difference \(\sigma_{33}-\sigma_{11}\), which disappears for the
  equilibrium value of \(c/a\). In order to apply this technique successfully,
  a number of technical sources of error must be brought under control.
  Firstly, the duration of the run must be long enough to reduce statistical 
  errors to within an acceptable level; secondly, the size of the simulation
  cell must be large enough; and thirdly, electronic \(k\)-point sampling must
  be adequate. The first two of the convergence questions can be addressed in
  detail using classical molecular dynamics simulations, and we have performed
  tests of this kind as described below. The question of \(k\)-point
  convergence will be discussed when we present our \textit{ab initio}
  molecular dynamics calculations in Sec.~\ref{ssec:aimd}.

  Although \textit{ab initio} molecular dynamics calculations fully include
  anharmonicity, they are very demanding in terms of computational resources,
  so that we can only perform these calculations at a very small number of
  state points. Our aim is to put limits on the possible size of anharmonic
  corrections to the stress, and hence to \(c/a\). Our \textit{ab initio}
  molecular dynamics results for \(\sigma_{33}-\sigma_{11}\) can be compared
  with our harmonic calculations by using Eqn.~\eqref{eq:stress} to calculate
  \(\sigma_{33}-\sigma_{11}\) in the harmonic approximation. Since
  anharmonicity is negligible at low temperatures, the harmonic and molecular
  dynamics calculations of the stress should be in close agreement under these
  conditions. This provides a further check both on our harmonic and molecular
  dynamics results.

  \subsection{Molecular dynamics simulations for an embedded-atom model}
  \label{ssec:belon}
  Our tests on statistical and system-size errors were performed using an
  embedded atom model (EAM) for Fe, similar to that used by \cite{baj:2000}.
  In the EAM, the total energy \(E_{\rm tot}\) of a system of \(N\) atoms
  is represented as the sum of two contributions. First, an empirical,
  repulsive pair potential, and secondly, a sum of embedding energies for each
  atom, representing quantum mechanical band-structure effects:
  \begin{equation}
    \begin{split}
      E_{\rm tot}&=E_{\rm pair}+E_{\rm embed}\\
      E_{\rm pair}&=\frac{1}{2}\sum_{i\ne j}\phi(|\mathbf{r}_i-\mathbf{r}_j|)\\
      E_{\rm embed}&=\sum_i f(\rho_i)\\
      \rho_i&=\sum_{j(\ne i)} \psi(|\mathbf{r}_i-\mathbf{r}_j|)
    \end{split}
  \end{equation}
  where \(\phi(r)\) and \(\psi(r)\) have the inverse power forms
  \begin{equation}
    \phi(r)=\epsilon \left(\frac{a}{r}\right)^n\quad \mbox{and}\quad
      \psi(r)=\left(\frac{a}{r}\right)^m\;,
  \end{equation}
  and \(f(\rho_i)=-\epsilon C(\rho_i)^{\frac{1}{2}}\).
  In practice, both \(\phi\) and \(\psi\) are cut off smoothly by making the
  transformation
  \begin{equation}
    \phi(r)\rightarrow \phi_{\rm SF}(r) = \left\{ \begin{array}{lll}
      \phi(r)+\beta+\gamma r                 &       & r < r_0 \\
      \alpha(r_1-r)^3                        & r_0 < & r < r_1 \\
      0                                      & r_1 < & r 
    \end{array} \right. \;,
  \end{equation}
  where \(r_0\) and \(r_1\) are cutoff radii, with \(\alpha\), \(\beta\) and
  \(\gamma\) chosen such that \(\phi\), \(\phi'\) and \(\phi''\) are
  continuous at \(r_0\). The form guarantees that these three functions are
  also continuous at \(r_1\). In the present calculations, we use
  \(r_1=7\)~\AA, \(r_0=0.9r_1\). The parameters have been modified somewhat
  from those given by \cite{baj:2000}, in order to reproduce better our
  \textit{ab initio} molecular dynamics simulations on solid and liquid Fe
  (\cite{apg:2001,apg:2002}). The parameters we user are: \(n=5.93\),
  \(m=4.788\), \(\epsilon=0.1662\)~eV, \(a=3.4714\)~\AA{} and \(C=16.55\).

\begin{figure}
    \vspace{4mm}
    \includegraphics[height=80mm]{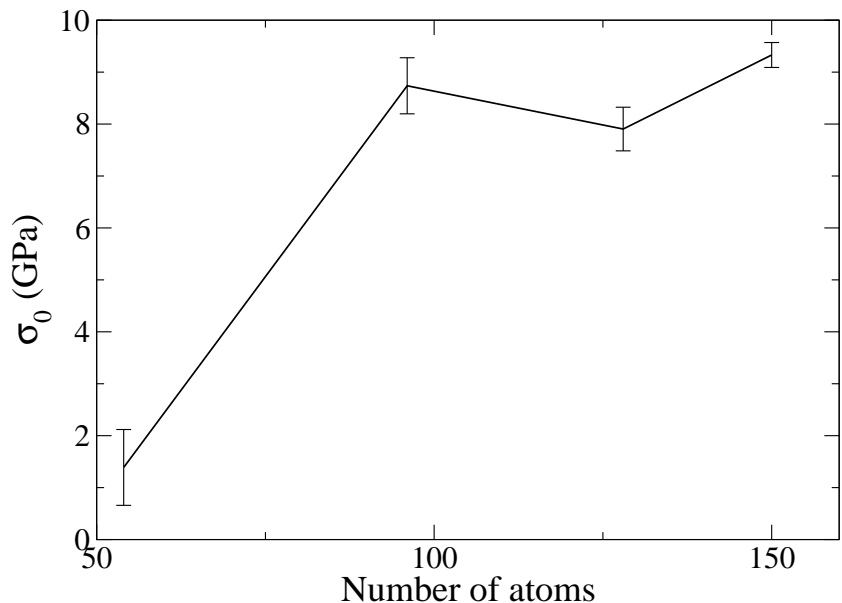}
    \caption{15~ps time-averaged calculations of
      \(\sigma_0=\sigma_{33}-\sigma_{11}\) for a variety of cell sizes within
      the embedded atom model.}
    \label{mdsize}
  \end{figure}

\begin{figure}
    \vspace{4mm}
    \includegraphics[height=80mm]{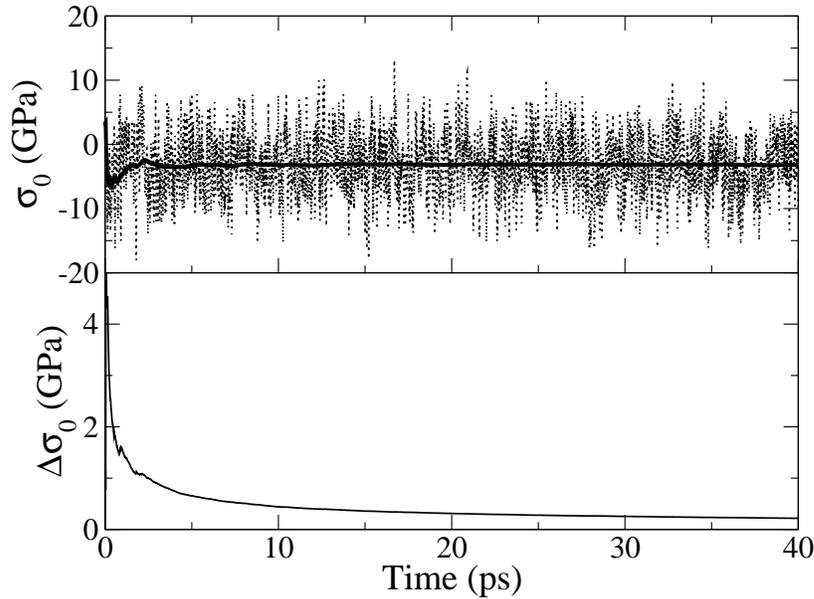}
    \caption{Instantaneous (dotted curve) and average (solid) value of
      \(\sigma_0=\sigma_{33}-\sigma_{11}\) (upper plot) and standard error
      (lower plot) for a 40~ps run for \(V=6.97\)~\AA\(^3/\mbox{atom}\),
      \(T=4000\)~K and \(c/a=1.65\) in the embedded atom model.}
    \label{run}
  \end{figure}

  In order to assess the length of run needed to reduce the statistical error
  on \(\sigma_{33}-\sigma_{11}\) to an acceptable level, we have performed
  molecular dynamics runs at several system sizes and at a number of different
  thermodynamic state points. An illustration of such a run is given in
  Fig.~\ref{run}. On the basis of these calculations we find that, in order to
  reduce the r.m.s. statistical errors to within 1~GPa, for example, we
  require a run of length 3--4~ps. Since the r.m.s. error is inversely
  proportional to
  the square root of the length of run, we may immediately infer the length of
  run necessary to obtain any required precision. From statistical mechanics,
  we expect that for a given length of run, the r.m.s. error on the stress
  components will decrease with increasing system size as \(N^{-\frac{1}{2}}\),
  where \(N\) is the number of atoms. We find that this is
  consistent with tests performed on system size, with r.m.s. errors on a
  1.5~ps average falling from 2~GPa for a 54-atom cell, to 1~GPa for 150
  atoms.
  To test size errors we performed runs on systems containing 54, 96, 128 and
  150 atoms at identical state points. Fig.~\ref{mdsize} shows the effect of
  system size on \(\sigma_{33}-\sigma_{11}\). Error bars show r.m.s.
  statistical errors as described above. We see that for a system of 96 atoms,
  the error due to size effects is within approximately 0.6~GPa.

  It follows from these tests that in order to calculate
  \(\sigma_{33}-\sigma_{11}\) to a tolerance needed to determine the
  equilibrium axial ratio to within 0.005, it should be enough to perform
  \textit{ab initio} molecular dynamics simulations of 1~ps on 96 atoms.

\begin{figure}
    \includegraphics[height=80mm]{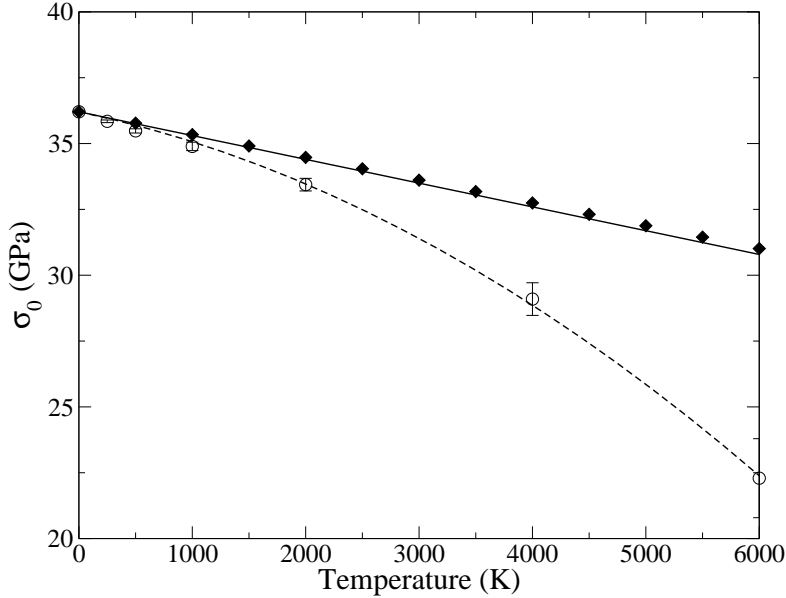}
    \caption{Molecular dynamics (open circles) and harmonic (diamonds)
      calculations of \(\sigma_0=\sigma_{33}-\sigma_{11}\) for a
      6.97~\AA\(^3\) cell in the embedded atom model, with \(c/a=1.70\). The
    dashed curve represents a second-order fit to the MD data, while the solid
    line is its tangent at 0~K. In agreement with thermodynamic theory, this
    coincides the harmonic calculations of the stress.}
    \label{anharm}
  \end{figure}

  As a further test of our techniques, we have compared molecular dynamics
  results for \(\sigma_{33}-\sigma_{11}\) with the predictions of
  harmonic theory for the embedded atom model. Fig.~\ref{anharm} presents the
  molecular dynamics and harmonic results for \(V=6.97\)~\AA\(^3\),
  \(c/a=1.65\) for a range of temperatures. Because of the gradual variation of
  \(\sigma_{33}-\sigma_{11}\) with temperature, we use longer runs of up
  to 5~ps. Cell sizes are converged withing the limits
  discussed above.
  We see that at low temperatures, the
  molecular dynamics results reproduce well the predictions of harmonic theory.
  This agreement provides additional confirmation that the statistical and
  system-size errors on the molecular dynamics calculations are very small.
  The deviation, rising to around 8.5~GPa at 6000~K, shows the high-temperature
  emergence of anharmonic effects.

  \subsection{\textit{Ab initio} molecular dynamics results}
  \label{ssec:aimd}
  
  We have performed \textit{ab initio} molecular dynamics simulations at an
  atomic volume of 6.97~\AA\(^3\), in which we have calculated \(\sigma_{33}
  - \sigma_{11}\). The molecular dynamics calculations are performed using
  exactly the same DFT methods as in the harmonic calculations. From the tests
  on the embedded atom model, we already have
  a good indication of the cell size and length of run necessary, but the
  question of
  \(k\)-points must also be addressed. To do this, we calculate \(\sigma_{33}
  - \sigma_{11}\) for several disordered configurations, selected from
  classical M.D. trajectories, both with \(\Gamma\)-point sampling and with
  larger numbers of \(k\)-points. Our tests are performed for a 96-atom
  system, with an atomic volume of 6.97~\AA\(^3\) and an axial ratio of 1.65.
  The molecular dynamics simulation from which the configurations were drawn
  was performed at 5000~K. We took 4 snapshots at intervals of 0.2 ps. For
  this state point, \(\sigma_{33}-\sigma_{11}\), with \(\Gamma\)-point sampling
  is equal to \(7.0\pm2\)~GPa. For a \(2\times2\times2\) \(k\)-point grid, it
  is equal to \(7.1\pm2.6\)~GPa. Similar tests were performed to ensure that
  size effects were consistent with the embedded atom model.
  
\begin{figure}
    \includegraphics[height=80mm]{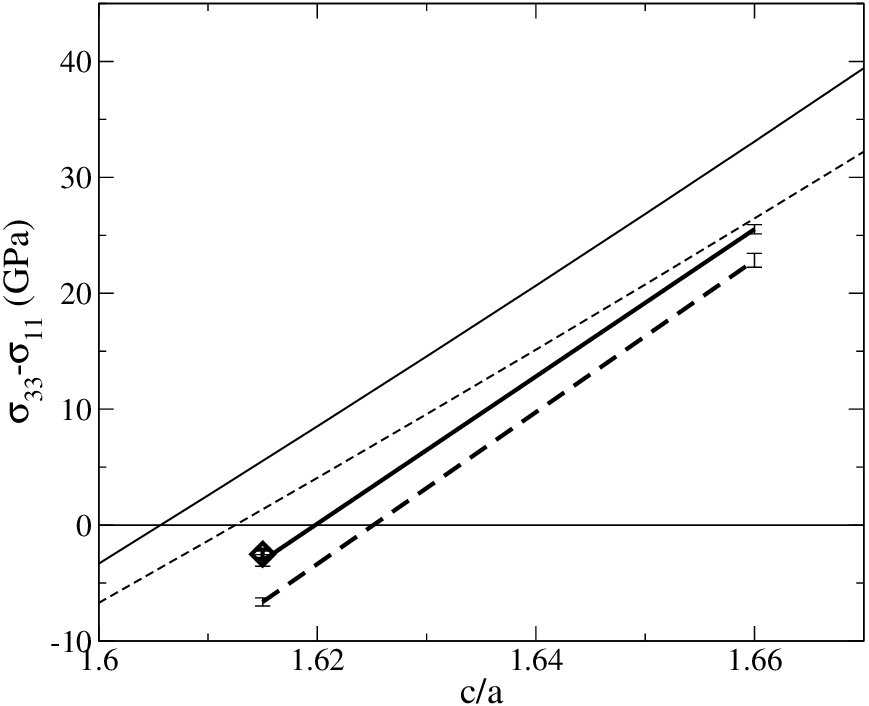}
    \caption{Comparison of \textit{ab initio} M.D. (heavy curves) and harmonic
      (light) results for the stress \(\sigma_{33}-\sigma_{11}\).  Solid lines
      are at 4000~K, dashed lines at 5000~K. The dotted line shows results for
      150-atom molecular dynamics at 4000~K. The diamond represents a
      96-atom run including 3\(p\) states explicitly in the valence set. All
      calculations were performed for an atomic volume of 6.97~\AA\(^3\). For
      comparison with earlier PIC calculations due to \cite{sscg:2001}, see
      Fig.~\ref{ca_temp}.}
    \label{md_harm}
  \end{figure}

  Our \textit{ab initio} calculations at 6.97~\AA\(^3\) were performed for
  \(c/a=1.615\) and 1.66, and for \(T=4000\)~K and 5000~K.  Fig.~\ref{md_harm}
  shows results for \(\sigma_{33}-\sigma_{11}\) along with harmonic
  predictions.
  We see that our molecular dynamics results closely resemble the
  predictions of harmonic theory, and vary in the same way with \(c/a\).
  However, even allowing for the error bars, there is an appreciable difference
  between the harmonic and M.D. results. One of the technical sources of error
  which could account for this discrepancy is system size effects. In order
  to check this, we have repeated the 5000~K runs using 150 atoms instead of
  96. The results of this test suggest that system-size errors could account
  for most of this discrepancy. We have also examined other technical sources
  of error, such as plane-wave cutoff and \(k\)-point sampling; however, as
  described above, there are not capable of producing such a discrepancy. It is
  possible in principle that some of this discrepancy can be attributed to
  anharmonic effects, but it is difficult to separate this from other effects
  with any certainty. Even allowing for the remaining uncertainties, it seems
  certain that anharmonic effects are not capable of shifting the equilibrium
  axial ratio by more than about 0.01.

  We have also used \textit{ab initio} M.D. to test another significant
  question. All our harmonic and M.D. calculations up to this point have
  treated 3\(p\) and 3\(s\) electrons as core states, but with the
  high-pressure response of these states treated via an empirical pair
  potential. In order to test the effect of this approximation, we have
  repeated the 96-atom M.D. calculations at 5000~K with 3\(p\) electrons
  explicitly included in the valence set. The resulting stress
  \(\sigma_{33}-\sigma_{11}\) agrees with our other results within our error
  bars. This indicates that the effects of this approximation are negligible
  for our present purposes.

  \section{Conclusions}
  \label{sec:conc}
  We have shown that it is rather straightforward to calculate harmonic free
  energies, and hence calculate the variation of \(c/a\) within the harmonic
  approximation. Because these calculations are computationally inexpensive, it
  is possible to apply them to a wide variety of thermodynamic state points.
  Since our earlier comparisons with experiment (\cite{mao_ea:2001}) show the
  reliability of our density functional methods for calculating the phonon
  frequencies for iron at high pressures, we expect our predictions of the
  temperature dependence of \(c/a\) to be reliable. We have also shown that
  direct \textit{ab initio} molecular dynamics simulation can be used to put
  limits on the possible size of anharmonic effects. These methods are rather
  expensive, but provide a means to perform ``spot checks'' on our results.
  The molecular dynamics calculations show that anharmonic corrections to the
  \(c/a\) ratio are small, even at temperatures near the melting point.

  The main scientific conclusion from the work is that \(c/a\) varies only
  rather weakly with temperature, and that its variation becomes weaker as
  pressures approach those of the inner core. Quantitatively, at a pressure of
  100~GPa, \(c/a\) increases from 1.585 at zero temperature, to 1.61 at the
  melting point, and at 300~GPa, from 1.59 at zero temperature, to 1.62 at the
  melting point. These results do not support the earlier predictions by
  \cite{sscg:2001} of a strong increase of \(c/a\) with temperature within the
  particle-in-cell approximation. Interestingly, the results we have found are
  very similar to our own predictions from the particle-in-cell approximation
  (\cite{gag:2003}). This implies that this approximation cannot be responsible
  for the large discrepancy between the present results and those of
  \cite{sscg:2001}. This finding of a rather weak increase of \(c/a\) with
  temperature is strongly supported by the recent diffraction experiments
  of \cite{mssmsh:2004}. Even allowing for the significant error bars on their
  \(c/a\) results, it seems clear that at 161~GPa, 2000~K, \(c/a\) is no bigger
  than about 1.61 at most. Our results would indicate a value of around 1.60
  under these conditions.

  Our results have implications for understanding the elastic anisotropy of the
  inner core. \cite{sscg:2001} predicted a reversal of the crystalline
  alignment required to explain the anisotropy of the inner core, due to a
  crossing of the \(c_{11}\) and \(c_{33}\) elastic moduli at
  approximately 1500~K. They attribute this to the strong
  temperature-dependence
  they find in \(c/a\). Since our results cast doubt on this strong
  temperature dependence, they also cast doubt on the proposed explanation for
  the elastic anisotropy. However, in order to resolve this question fully, we
  need to make \textit{ab initio} predictions of the elastic moduli, that are
  free of statistical mechanical approximations. The methods we have presented
  here should be capable of achieving this, and we hope to be able to present
  such results in due course.

  \ack
  The work of CMSG is supported by a NERC studentship, and the work of DA by a
  Royal Society fellowship and by an award from the Leverhulme Trust. The
  calculations were performed using UCL facilities, supported by a JREI grant
  and by SRIF funding. The authors thank G.~D.~Price and L.~Vo\v{c}adlo for
  useful discussions.

  \bibliography{all_papers}

\begin{thebibliography}{46}
\expandafter\ifx\csname natexlab\endcsname\relax\def\natexlab#1{#1}\fi
\expandafter\ifx\csname url\endcsname\relax
  \def\url#1{\texttt{#1}}\fi
\expandafter\ifx\csname urlprefix\endcsname\relax\def\urlprefix{URL }\fi

\bibitem[{Alf\'e(1998)}]{alfe_phon}
Alf\'e, D., 1998. Program available at
  http://chianti.geol.ucl.ac.uk/\(\sim\)dario.

\bibitem[{Alf\`{e} et~al.(2002{\natexlab{a}})Alf\`{e}, Gillan, and
  Price}]{agp:2002}
Alf\`{e}, D., Gillan, M.~J., Price, G.~D., 2002{\natexlab{a}}. Ab initio
  chemical potentials of solid and liquid solutions and the chemistry of the
  \uppercase{E}arth's core. J. Chem. Phys. 116, 7127.

\bibitem[{Alf{\`e} et~al.(2000)Alf{\`e}, Kresse, and Gillan}]{akg:2000}
Alf{\`e}, D., Kresse, G., Gillan, M.~J., 2000. Structure and dynamics of liquid
  iron under \uppercase{E}arth's core conditions. Phys. Rev. B 61, 132.

\bibitem[{Alf\`{e} et~al.(2001)Alf\`{e}, Price, and Gillan}]{apg:2001}
Alf\`{e}, D., Price, G.~D., Gillan, M.~J., 2001. Thermodynamics of
  hexagonal-close-packed iron under \uppercase{E}arth's core conditions. Phys.
  Rev. B 64, 045123.

\bibitem[{Alf\`{e} et~al.(2002{\natexlab{b}})Alf\`{e}, Price, and
  Gillan}]{apg:2002}
Alf\`{e}, D., Price, G.~D., Gillan, M.~J., 2002{\natexlab{b}}. Iron under
  \uppercase{E}arth's core conditions: Liquid-state thermodynamics and
  high-pressure melting curve from ab initio calculations. Phys. Rev. B 65,
  165118.

\bibitem[{Andersen(1980)}]{andersen:1980}
Andersen, H.~C., 1980. Molecular dynamics simulations at constant pressure
  and/or temperature. J. Chem. Phys. 72, 2384.

\bibitem[{Beghein and Trampert(2003)}]{bt:2003}
Beghein, C., Trampert, J., 2003. Robust normal mode constraints on inner-core
  anisotropy from model space search. Science 299, 552.

\bibitem[{Belonoshko et~al.(2000)Belonoshko, Ahuja, and Johansson}]{baj:2000}
Belonoshko, A.~B., Ahuja, R., Johansson, B., 2000. Quasi-\textit{ab initio}
  molecular dynamic study of \uppercase{F}e melting. Phys. Rev. Lett. 84, 3638.

\bibitem[{Belonoshko et~al.(2003)Belonoshko, Ahuja, and Johansson}]{baj:2003}
Belonoshko, A.~B., Ahuja, R., Johansson, B., 2003. Stability of the
  body-centred-cubic phase of iron in the \uppercase{E}arth's inner core.
  Nature 424, 1032.

\bibitem[{Bl{\"o}chl(1994)}]{blochl:1994}
Bl{\"o}chl, P.~E., 1994. Projector augmented-wave method. Phys. Rev. B 50, 953.

\bibitem[{Cowley et~al.(1990)Cowley, Gross, Gong, and Horton}]{c_ea:1990}
Cowley, E.~R., Gross, J., Gong, Z.~X., Horton, G.~K., 1990. Cell-cluster and
  self-consistent calculations for a model sodium-chloride crystal. Phys. Rev.
  B 42, 3135.

\bibitem[{Creager(1992)}]{creager:1992}
Creager, K.~C., 1992. Anisotropy of the inner core from differential
  travel-times of the phases pkp and pkikp. Nature 356, 309.

\bibitem[{Gannarelli et~al.(2003)Gannarelli, Alf{\`e}, and Gillan}]{gag:2003}
Gannarelli, C. M.~S., Alf{\`e}, D., Gillan, M.~J., 2003. The particle-in-cell
  model for ab initio thermodynamics: implications for the elastic anisotropy
  of the \uppercase{E}arth's inner core. Phys. Earth Planet. Inter 139, 243.

\bibitem[{Hirshfelder et~al.(1954)Hirshfelder, Curtiss, and Bird}]{hcb:1954}
Hirshfelder, J.~O., Curtiss, C.~F., Bird, R.~B., 1954. Molecular Theory of
  Gases and Liquids. John Wiley and Sons, Inc, New York.

\bibitem[{Hohenberg and Kohn(1964)}]{hk:1964}
Hohenberg, P., Kohn, W., 1964. Inhomogeneous electron gas. Phys. Rev. 136,
  B864.

\bibitem[{Holt et~al.(1970)Holt, Hoover, Gray, and Shortle}]{h_ea:1970}
Holt, A.~C., Hoover, W.~G., Gray, S.~G., Shortle, D.~R., 1970. Comparison of
  the lattice-dynamics and cell-model approximations with
  \uppercase{M}onte-\uppercase{C}arlo thermodynamic properties. Physica 49, 61.

\bibitem[{Holt and Ross(1970)}]{hr:1970}
Holt, A.~C., Ross, M., 1970. Calculations of the \uppercase{G}r{\"u}neisen
  parameter of some models of the solid. Phys. Rev. B 1, 2700.

\bibitem[{Ishii and Dziewonski(2002)}]{id:2002}
Ishii, M., Dziewonski, A.~M., 2002. The innermost inner core of the
  \uppercase{E}arth: Evidence for a change in anisotropic behaviour at the
  radius of about 300~km. Proc. Natl. Acad. Sci USA 99, 14026.

\bibitem[{Jones and Gunnarsson(1989)}]{jg:1989}
Jones, R.~O., Gunnarsson, O., 1989. The density functional formalism, its
  applications and prospects. Rev. Mod. Phys. 61, 689.

\bibitem[{Kohn and Sham(1965)}]{ks:1965}
Kohn, W., Sham, L., 1965. Self-consistent equations including exchange and
  correlation effects. Phys. Rev. 140, A1133.

\bibitem[{Kresse and Furthm{\"u}ller(1996{\natexlab{a}})}]{kf_02:1996}
Kresse, G., Furthm{\"u}ller, J., 1996{\natexlab{a}}. Efficiency of \textit{ab
  initio} total energy calculations for metals and semiconductors using a
  plane-wave basis set. Comput. Mater. Sci. 6, 15.

\bibitem[{Kresse and Furthm{\"u}ller(1996{\natexlab{b}})}]{kf_01:1996}
Kresse, G., Furthm{\"u}ller, J., 1996{\natexlab{b}}. Efficient iterative
  schemes for \textit{ab initio} total-energy calculations using a plane-wave
  basis set. Phys. Rev. B 54, 11169.

\bibitem[{Kresse et~al.(1995)Kresse, Furthm{\"u}ller, and Hafner}]{kfh:1995}
Kresse, G., Furthm{\"u}ller, J., Hafner, J., 1995. Force constant approach to
  phonon dispersion relations of diamond and graphite. Europhys. Lett. 32, 729.

\bibitem[{Kresse and Joubert(1999)}]{kj:1999}
Kresse, G., Joubert, D., 1999. From ultrasoft pseudopotentials to the projector
  augmented-wave method. Phys. Rev. B 59, 1758.

\bibitem[{Laio et~al.(2000)Laio, Bernard, Chiarotti, Scandolo, and
  Tosatti}]{lbcst:2000}
Laio, A., Bernard, S., Chiarotti, G.~L., Scandolo, S., Tosatti, E., 2000.
  Physics of iron at \uppercase{E}arth's core conditions. Science 287, 1027.

\bibitem[{Ma et~al.(2004)Ma, Somayazulu, Shen, Mao, Shu, and
  Hemley}]{mssmsh:2004}
Ma, Y., Somayazulu, M., Shen, G., Mao, H.~K., Shu, J., Hemley, R., 2004. In
  situ x-ray diffraction studies of iron to \uppercase{E}arth-core conditions.
  Phys. Earth Planet. Inter 143-144, 455.

\bibitem[{Mao et~al.(2001)Mao, Xu, Struzhkin, Shu, Hemley, Sturhahn, Hu, Alp,
  Vo{\v c}adlo, Alf{\`e}, Price, Gillan, Schwoerer-B{\"o}hning, H{\"a}usermann,
  Eng, Shen, Giefers, L{\"u}bbers, and Wortmann}]{mao_ea:2001}
Mao, H.~K., Xu, J., Struzhkin, V.~V., Shu, J., Hemley, R.~J., Sturhahn, W., Hu,
  M.~Y., Alp, E.~E., Vo{\v c}adlo, L., Alf{\`e}, D., Price, G.~D., Gillan,
  M.~J., Schwoerer-B{\"o}hning, M., H{\"a}usermann, D., Eng, P., Shen, G.,
  Giefers, H., L{\"u}bbers, R., Wortmann, G., 2001. Phonon density of states of
  iron up to 153 gigapascals. Science 292, 914.

\bibitem[{Mermin(1965)}]{mermin:1965}
Mermin, N.~D., 1965. Thermal properties of the inhomogeneous electron gas.
  Phys. Rev. 137, A1441.

\bibitem[{Oganov et~al.(2001)Oganov, Brodholt, and Price}]{obp:2001}
Oganov, A.~R., Brodholt, J.~P., Price, G.~D., 2001. The elastic constants of
  \uppercase{M}g\uppercase{S}i\uppercase{O}\(_3\) perovskite at pressures and
  temperatures of the \uppercase{E}arth's mantle. Nature 411, 934.

\bibitem[{Poirier(1994)}]{poirier:1994}
Poirier, J.~P., 1994. Light elements in the \uppercase{E}arth's outer core: A
  critical review. Phys. Earth Planet. Inter 85, 319.

\bibitem[{Ree and Holt(1973)}]{rh:1973}
Ree, F.~H., Holt, A.~C., 1973. Thermodynamic properties of the alkali-halide
  crystals. Phys. Rev. B 8, 826.

\bibitem[{S{\"o}derlind et~al.(1996)S{\"o}derlind, Moriarty, and
  Willis}]{smw:1996}
S{\"o}derlind, P., Moriarty, J.~A., Willis, J.~M., 1996. First-principles
  theory of iron up to earth-core pressures: Structural, vibrational, and
  elastic properties. Phys. Rev. B 53, 14063.

\bibitem[{Song and Helmberger(1998)}]{sh:1998}
Song, X.~D., Helmberger, D.~V., 1998. Seismic evidence for an inner core
  transition zone. Science 282, 924.

\bibitem[{Steinle-Neumann et~al.(1999)Steinle-Neumann, Stixrude, and
  Cohen}]{ssc:1999}
Steinle-Neumann, G., Stixrude, L., Cohen, R.~E., 1999. First-principles elastic
  constants for the hcp transition metals \uppercase{F}e, \uppercase{C}o, and
  \uppercase{R}e at high pressure. Phys. Rev. B 60, 791.

\bibitem[{Steinle-Neumann et~al.(2002)Steinle-Neumann, Stixrude, and
  Cohen}]{ssc:2002}
Steinle-Neumann, G., Stixrude, L., Cohen, R.~E., 2002. Physical properties of
  iron in the inner core. In: Dehant, V., Creager, K., Zatman, S., Karato,
  S.-I. (Eds.), Core structure, dynamics and rotation. American Geophysical
  Union, Washington, DC, pp. 137--161.

\bibitem[{Steinle-Neumann et~al.(2001)Steinle-Neumann, Stixrude, Cohen, and
  G{\"u}lseren}]{sscg:2001}
Steinle-Neumann, G., Stixrude, L., Cohen, R.~E., G{\"u}lseren, O., 2001.
  Elasticity of iron at the temperature of the \uppercase{E}arth's inner core.
  Nature 413, 57.

\bibitem[{Stixrude and Cohen(1995)}]{sc:1995}
Stixrude, L., Cohen, R.~E., 1995. High-pressure elasticity of iron and
  anisotropy of \uppercase{E}arth's inner-core. Science 267, 1972.

\bibitem[{Stixrude et~al.(1994)Stixrude, Cohen, and Singh}]{scs:1994}
Stixrude, L., Cohen, R.~E., Singh, D.~J., 1994. Iron at high pressure:
  Linearized-augmented-plane-wave computations in the generalized-gradient
  approximation. Phys. Rev. B 50, 6442.

\bibitem[{Stixrude et~al.(1997)Stixrude, Wasserman, and Cohen}]{swc:1997}
Stixrude, L., Wasserman, E., Cohen, R.~E., 1997. Composition and temperature of
  \uppercase{E}arth's inner core. J. Geophys. Res., [Space Phys] 102, 24729.

\bibitem[{Tromp(1993)}]{tromp:1993}
Tromp, J., 1993. Support for anisotropy of the \uppercase{E}arth's inner-core
  from free oscillations. Nature 366, 678.

\bibitem[{Vo{\v c}adlo et~al.(2003)Vo{\v c}adlo, Alf{\`e}, Gillan, Wood,
  Brodholt, and Price}]{vagwbp:2003}
Vo{\v c}adlo, L., Alf{\`e}, D., Gillan, M.~J., Wood, I.~G., Brodholt, J.~P.,
  Price, G.~D., 2003. Possible thermal and chemical stabilisation of
  body-centred-cubic iron in the \uppercase{E}arth's core. Nature 424, 536.

\bibitem[{Vocadlo et~al.(2000)Vocadlo, Brodholt, Alfe, Gillan, and
  Price}]{vbagp:2000}
Vocadlo, L., Brodholt, J., Alfe, D., Gillan, M.~J., Price, G.~D., 2000. Ab
  initio free energy calculations on the polymorphs of iron at core conditions.
  Phys. Earth Planet. Inter 117, 123.

\bibitem[{Vo{\v c}adlo et~al.(1997)Vo{\v c}adlo, de~Wijs, Kresse, Gillan, and
  Price}]{vwkgp:1997}
Vo{\v c}adlo, L., de~Wijs, G.~A., Kresse, G., Gillan, M.~J., Price, G.~D.,
  1997. First principles calculations on crystalline and liquid iron at
  \uppercase{E}arth's core conditions. Faraday Discuss. 106, 205.

\bibitem[{Wang and Perdew(1991)}]{wp:1991}
Wang, Y., Perdew, J.~P., 1991. Correlation hole of the spin-polarized electron
  gas, with exact small-wave-vector and high-density scaling. Phys. Rev. B 44,
  13298.

\bibitem[{Wasserman et~al.(1996)Wasserman, Stixrude, and Cohen}]{wsc:1996}
Wasserman, E., Stixrude, L., Cohen, R.~E., 1996. Thermal properties of iron at
  high pressures and temperatures. Phys. Rev. B 53, 8296.

\bibitem[{Westra and Cowley(1975)}]{wc:1975}
Westra, K., Cowley, E.~R., 1975. Cell-cluster expansion for an anharmonic
  solid. Phys. Rev. B 11, 4008.

\end{thebibliography}
\end{document}